\documentclass[aps,prd,12point,twocolumn,nofootinbib,showpacs,superscriptaddress]{revtex4-1}

\usepackage{amssymb}
\usepackage{graphicx}
\usepackage{amsmath, amsthm}
\usepackage{epstopdf}
\usepackage{hyperref}
\usepackage{enumerate}

\newcommand{\be}{\begin{equation}}
\newcommand{\ee}{\end{equation}}

\begin{document}

\title{Black holes and wormholes subject to conformal 
mappings} 

\author{Valerio Faraoni} 
\email[]{vfaraoni@ubishops.ca}
\affiliation{Physics Department,  
Bishop's University, Sherbrooke, Qu\'ebec, Canada J1M~1Z7
}

\author{Angus Prain}
\email[]{aprain@ubishops.ca}
\affiliation{Physics Department,  
Bishop's University, 
Sherbrooke, Qu\'ebec, Canada J1M~1Z7
}

\author{Andres F. Zambrano Moreno}
\email[]{andres.zambranomoren@ucalgary.ca}
\affiliation{Physics Department,  
Bishop's University, 
Sherbrooke, Qu\'ebec, Canada J1M~1Z7
}
\affiliation{Department of Physics and Astronomy, 
University of Calgary, Alberta, Canada T2N~1N4}


\begin{abstract} 
Solutions of the field equations of theories of 
gravity which 
admit distinct conformal frame representations can look 
very different in these frames. We show that Brans class~IV 
solutions describe wormholes in the Jordan frame (in a 
certain parameter range) but correspond to horizonless 
geometries in the Einstein frame. 
The reasons for such a change of behaviour 
under conformal mappings are elucidated in general, 
using Brans~IV solutions as an example.
\end{abstract}



\maketitle

\section{Introduction} 
\label{sec:1}

Conformal transformations are widely used in General 
Relativity \cite{Waldbook} and especially in alternative 
theories of gravity \cite{BD61, FujiiMaeda, mybook, 
CapozzielloFaraonibook}. 
Scalar-tensor and $f(R)$ gravity (which can be 
reduced to the former \cite{reviews}) can be represented in 
infinitely many conformal frames, of which the Jordan and 
the Einstein frame are commonly used \cite{FujiiMaeda, 
mybook}. By conformally 
transforming a black 
hole, often the result is not another black hole 
geometry. Other times, a horizonless geometry (possibly 
containing a naked singularity) is conformally transformed 
to a black hole solution of the field equations or even to 
a wormhole and, {\em vice-versa}, a black hole  
can be transformed into a naked singularity. Although this 
phenomenology has been noted occasionally in the 
literature ({\em e.g.}, \cite{noted1, noted2, noted3, 
Alex}), to the best of our knowledge it has not been 
investigated. Here we fill this gap  by 
tracing the cause of the disappearance of horizons and the 
appearance of new ones under the action of conformal 
transformations. For concreteness, we apply our discussion 
to scalar-tensor gravity, but the context is more general 
and our results can be applied to any situation in which a 
spacetime metric is conformally transformed, including in 
General Relativity. For 
simplicity, however, we consider spherically symmetric 
metrics and we provide an example which is, in 
addition, static. 

In the course of our discussion we 
elucidate (Sec.~\ref{sec:2}) the nature of the well known 
Brans class~IV solutions of Brans-Dicke gravity, which 
constitute our example for the general problem. For a 
certain range of parameters, they are found to represent 
asymptotically flat wormholes with a horizon, while for 
other parameter ranges they are horizonless geometries 
hosting naked singularities. 
This radical change in geometry is not limited to this 
example but occurs frequently when conformal mappings of 
spacetime metrics are involved, not only in scalar-tensor 
gravity but also in General Relativity. The reasons for 
such a change in geometry are investigated and 
clarified in Sec.~\ref{sec:3} for general situations.

We remind the reader that the (Jordan frame) action of 
scalar-tensor gravity is 
\begin{eqnarray}
S_\text{ST} &=& \int d^4x \sqrt{-g} \left[ 
\frac{1}{16\pi} \left( \phi {\cal R}  
-\frac{\omega(\phi)}{\phi} \, \nabla^a \phi \nabla_a \phi 
\right) -V(\phi) \right. \nonumber\\
&&\nonumber\\
&\, &\left.  +{\cal 
L}_\text{matter} \right]\,,
\end{eqnarray}
where ${\cal R}$ is the Ricci curvature of spacetime, $g$ is the 
determinant of the spacetime metric $g_{ab}$, and $ \phi$ is the 
Brans-Dicke-like scalar field, while $V(\phi)$ is a scalar field 
potential.  A conformal transformation of the 
metric $g_{ab} \rightarrow \tilde{g}_{ab} =\Omega^2 g_{ab}$ with  
conformal factor $\Omega=\sqrt{\phi}$,  together with the 
scalar field redefinition
\be
d\tilde{\phi} = \sqrt{ \frac{|2\omega (\phi) +3|}{16 \pi}} 
\, \frac{ d\phi}{\phi} 
\ee
recasts the theory in its 
Einstein frame form 
\begin{eqnarray}
S_\text{ST} &=& 
\int d^4 x \sqrt{-\tilde{g}} \left[ 
\frac{\cal R}{16\pi} -\frac{1}{2} \, \tilde{g}^{ab} 
\nabla_a \tilde{\phi}  \nabla_b \tilde{\phi}  -U\left( 
\tilde{\phi}\right) \right.\nonumber\\
&&\nonumber\\
&\, & \left. +\frac{ {\cal 
L}_\text{matter} }{ 
\left( \phi( \tilde{\phi}) \right)^2 } 
\right]
\end{eqnarray}
in which the gravity sector looks like General Relativity 
and the new scalar field $\tilde{\phi}$ couples minimally 
to gravity (but non-minimally to matter) and has canonical 
kinetic energy density and 
potential 
\be
U\left( \tilde{\phi} \right)=\frac{ V \left( 
\phi( \tilde{\phi}) \right)}{
\left( \phi( \tilde{\phi}) \right)^2} \,.
\ee

We use units in which Newton's constant $G$ and the speed 
of light $c$ are unity, the metric signature is $-+++$, and 
we follow the notations and conventions of 
Ref.~\cite{Waldbook}.

\section{Brans class~IV solutions}
\label{sec:2}

Brans class~IV spacetimes \cite{Brans62} 
form an often quoted class of static and spherically 
symmetric solutions of vacuum Brans-Dicke 
theory 
(without scalar field potential). 
Class~III and class~IV solutions \cite{Brans62} can be obtained 
from each 
other 
by means of a mathematical transformation involving the radial 
coordinate and the parameters 
\cite{BhadraNandi2001, BhadraSarkarGRG2005}. Although the discovery of these solutions was made only one year after the introduction of Brans-Dicke theory \cite{Brans62}, and often cited, they are still not well uderstood.

The  line 
element and Brans-Dicke scalar field in isotropic 
coordinates in the Jordan frame are \cite{Brans62}
\be
ds^2= -\mbox{e}^{2\alpha_0 -\frac{2}{B_0 \varrho} } dt^2 
+\mbox{e}^{2\beta_0 +\frac{2(C_0+1)}{B_0\varrho}} \left( 
d\varrho^2 +\varrho^2 d\Omega_{(2)}^2\right) \,, 
\label{BransIV}
\ee
\be
\phi( \varrho) = \phi_0 \, \mbox{e}^{-\frac{C_0}{B_0 
\varrho}} 
\,,
\label{BransIVscalar}
\ee
where $d\Omega_{(2)}^2=d\theta^2 +\sin^2 \theta \, 
d\varphi^2$ is the line element on the unit 2-sphere and 
\be
C_0= \frac{-1\pm \sqrt{-2\omega-3}}{\omega+2} 
\ee
(which requires that $2\omega+3 \leq 0$), and where 
$\alpha_0 , \beta_0, B_0,C_0$, and $\phi_0$ are 
constants. The factors $\mbox{e}^{2\alpha_0}$ 
and $\mbox{e}^{2\beta_0}$ can be absorbed by rescaling the 
coordinates $t$ and $\varrho$, respectively, and in the 
following they are  dropped. 

A notable feature of Brans class~IV solutions is that, as 
the Brans-Dicke parameter $\omega \rightarrow -\infty$, they
do not reduce to the corresponding vacuum solution of 
General Relativity, the Schwarzschild solution, but to 
\be
ds^2 \approx -\mbox{e}^{-\frac{2}{B_0 \varrho} } dt^2 
+\mbox{e}^{\frac{2}{B_0\varrho}} \left( d\varrho^2 
+\varrho^2 d\Omega_{(2)}^2\right) \,, 
\ee
\be
\phi( \varrho) \approx \phi_0 \, .
\ee
This anomalous behaviour is related to the fact that the 
scalar 
$\phi$ does not have the usual 
\cite{Weinberg} asymptotic behaviour 
$\phi =\phi_0 
+ \frac{\phi_1}{\omega} +\frac{\phi_2}{\omega^2} + \, ...$ 
as $|\omega| \rightarrow +\infty$, but rather follows 
$\phi=\phi_0 +\mbox{O} \left( 1/\sqrt{|\omega |}  \right) 
$. 
It is well known 
that Brans class~I solutions 
exhibit the same phenomenon  \cite{RomeroBarros, 
BanerjeeSen, VFlimit, BhadraNandilimit}, so the lack of  
a limit to General Relativity for class~IV solutions is not 
a surprise.

\begin{figure}
\includegraphics[scale=0.4]{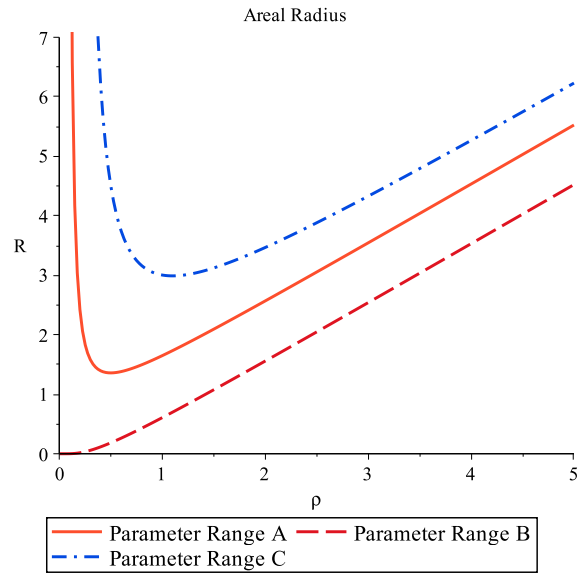}
\caption{The areal radius $R(\varrho)$ for the three ranges 
of values of the parameters $B_0$ 
and $C_0$ considered in the text. We use  
$(B_0,C_0)=(1,-1/2)$, $(-1,-1/2)$ and $(-1,1.9)$ for 
parameter ranges A, B and C, respectively. 
\label{F:ranges}}
\end{figure}

The areal radius of the Brans class~IV  
geometry~(\ref{BransIV}) is  
\be\label{arealradius}
R(\varrho)= \mbox{e}^{ \frac{C_0+1}{B_0\varrho} } 
\, \varrho \,.
\ee
Let us see how 2-spheres of symmetry behave as the 
isotropic radius $\varrho$ varies. We have
\be
\frac{dR}{d\varrho}= \mbox{e}^{ \frac{C_0+1}{B_0\varrho} } 
\left(1-\frac{\varrho_0}{\varrho} \right)
\ee
where
\be\label{rho0}
\varrho_0 \equiv \frac{C_0+1}{B_0} \,.
\ee
Assuming $\omega<-3/2$ and $B_0 \neq 0$, we consider 
separately three possible parameter ranges which we present in Fig.~\ref{F:ranges}.

\subsection{Parameter range $B_0>0, \, C_0 >-1$}

In this case it is $dR/d\varrho \geq 0 $ for $\varrho \geq  
\varrho_0$ and this derivative is negative otherwise: the 
areal radius $R(\varrho) $ diverges as $\varrho \rightarrow 
0^{+}$, decreases as $\varrho $ increases, assumes a global 
minimum $R_\text{min}=\mbox{e} \varrho_0 $ at 
$\varrho_0$, and increases to $+ \infty$ as $\varrho 
\rightarrow + \infty$. There are, therefore, two branches 
of 
the areal radius $R$ as $\varrho$ varies in the range $0 
\leq 
\varrho < +\infty$. In order to understand what happens at 
$\varrho_0$, we rewrite the line element~(\ref{BransIV}) 
using 
the areal radius instead of the coordinate $\varrho$, which 
gives straightforwardly
\be
ds^2= -\mbox{e}^{ -\frac{2}{B_0 \varrho} } dt^2 
+\frac{dR^2}{\left( 1-\varrho_0/\varrho \right)^2} +R^2 
d\Omega_{(2)}^2 \,.
\ee
The apparent/trapping horizons (if any exist) of any 
spherically symmetric metric are located at the roots of 
the equation $\nabla^c R\nabla_c R=0$ 
\cite{NielsenVisser, AbreuVisser, VFhorizons} which, in 
these coordinates, correspond 
to $g^{RR}=0$ or $\left( 1-\varrho_0/\varrho \right)^2=0$, 
which 
has $\varrho_0$ as a double root. Here, $g^{RR}$ does not change 
sign at the minimum $R_\text{min}$.  In general wherever $g^{RR}$ changes sign, so that  $\partial g^{RR}/\partial R 
\neq 0$ there and $g^{RR}$ has a single 
root, there exists a  
black hole horizon. In this black hole case, the isotropic radius $\varrho >0$  
corresponds to a double 
covering of the exterior region \cite{Buchdhal}. Therefore, 
for the range of parameters we are considering here, 
$\varrho=\varrho_0$ 
corresponds to an horizon, but not to a black 
hole horizon: the line element~(\ref{BransIV}) corresponds 
to a wormhole with the two throats joining at the horizon. 
This conclusion is 
similar to that reached in 
Ref.~\cite{Vanzoetal2012} for the Campanelli-Lousto 
spacetimes \cite{CampanelliLousto}, another class of 
static and spherically symmetric solutions 
of Brans-Dicke theory which are often (erroneously) 
referred to as black holes.
 
The Ricci, Ricci squared, and Kretschmann scalars are
\begin{eqnarray}
{\cal R} &=& \frac{2}{R^2} \left[ 
1-\left( 1-\frac{\varrho_0}{\varrho} \right)^2\right]  
\,,\label{Ricciscalar}\\
&&\nonumber\\
R_{ab}R^{ab} &=& \frac{2}{R^4} 
\left[ 1-\left( 1-\frac{\varrho_0}{\varrho} 
\right)^2\right]^2 
\,,\label{Riccisquared}\\
&&\nonumber\\
R_{abcd}R^{abcd} &=& \frac{4}{R^4} 
\left[ 1-\left( 1-\frac{\varrho_0}{\varrho} 
\right)^2\right]^2 
\,,\label{Riemannsquared}
\end{eqnarray}
respectively. These scalars do not diverge since $R$ stops 
at its minimum $R_\text{min}$ and does not reach zero 
value.

\subsection{Parameter range $B_0<0, C_0>-1$}
 
For the parameter values $B_0<0, \, C_0 >-1$, it is 
$\varrho_0 \equiv 
\frac{C_0+1}{B_0} <0$ and $dR/d\varrho >0 $ for any 
positive 
value 
of $\varrho$, hence  the areal 
radius has no positive minimum. 
The function $R(\varrho)$ increases monotonically from 
$R(0)=0$ to positive infinity as $\varrho$ increases. In 
this case there are no horizons.
All the scalar invariants 
(\ref{Ricciscalar})-(\ref{Riemannsquared}) diverge as 
$R\rightarrow 0$ (corresponding to $\varrho \rightarrow 
0$). This spacetime harbours a naked singularity at $R=0$.

\subsection{Parameter range $B_0<0, \, C_0<-1$}

For this value of the parameters it is again 
$\varrho_0=\left| 
\frac{C_0+1}{B_0} \right|>0$ and  the function $R(\varrho)$ 
is again decreasing for $0 \leq \varrho <\varrho_0$, 
minimum 
at 
$\varrho_0$, and increasing for $\varrho>\varrho_0$. We 
have 
again 
two wormhole throats joining at an horizon located at 
$R_\text{min}$ and no spacetime singularities are present.

\subsection{Einstein frame Brans~IV metric} 

Let us consider now, for all values of the parameters $B_0$ 
and $C_0$, the Einstein frame version of Brans-Dicke theory. 
By performing the usual conformal rescaling of the 
metric to the Einstein frame
\be
g_{ab} \rightarrow \tilde{g}_{ab}=\Omega^2 g_{ab} \,, 
\;\;\;\;\; \Omega=\sqrt{\phi} 
\ee
and the nonlinear scalar field redefinition
\be
\phi \rightarrow \tilde{\phi}=\sqrt{ 
\frac{|2\omega+3|}{16\pi}} \, \ln \left( 
\frac{\phi}{\phi_*} \right)
\ee
where $\phi_*$ is a constant, the theory is recast in the 
form of Einstein gravity with a minimally coupled scalar 
field. Dropping irrelevant constants, one obtains the 
Einstein frame line element and scalar field
\be
d\tilde{s}^2= -\mbox{e}^{- \frac{ (C_0+2)}{B_0\varrho}}  
dt^2 
+ \mbox{e}^{\frac{C_0+2}{B_0\varrho}} \left( d\varrho^2 
+\varrho^2 
d\Omega_{(2)}^2 \right) \,, \label{E:non_tilded}
\ee
\be
\tilde{\phi}= \phi_1-\frac{\phi_2}{\varrho} \,,
\ee
where
\begin{eqnarray}
\phi_1 &=& \sqrt{ \frac{ |2\omega+3|}{16\pi}} \, \ln \left( 
\frac{\phi}{\phi_*} \right) \,,\\
&&\nonumber\\
\phi_2 &=& -\sqrt{ \frac{ |2\omega+3|}{16\pi}} 
\,\frac{C_0}{B_0} \,.
\end{eqnarray}

\begin{figure}
\includegraphics[scale=0.4]{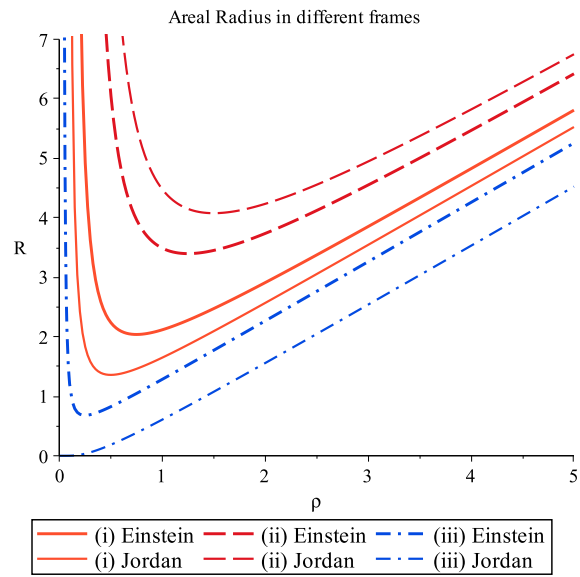}
\caption{Einstein frame and Jordan frame areal radius 
for the three scenarios described in the text. 
\label{F:scenarios}}
\end{figure}

Now the areal radius is not given by~(\ref{arealradius}) 
but by  
\be
\tilde{R}=\mbox{e}^{\frac{C_0+2}{2B_0\varrho}} \, \varrho
\ee
which differs from its Jordan frame counterpart in the 
numerical coefficient of $1/(B_0\varrho)$ in the exponent. The same 
analysis 
performed in the Jordan frame can be repeated here but now 
the critical isotropic radius is not $\varrho_0$ but 
\be\label{rho1}
\tilde{\varrho}_0  \equiv \frac{C_0+2}{2B_0} = 
\frac{\varrho_0}{2}+\frac{1}{2B_0} \,.
\ee  
Assuming $B_0>0$, one immediately sees that 
\begin{enumerate}[(i)]
\item $\tilde{\varrho}_0>\varrho_0 $ if $-1<C_0<0$;
\item $\tilde{\varrho}_0<\varrho_0 $ if $C_0>0$;
\item but $ \tilde{\varrho}_0>0$ and $\varrho_0<0 $ for 
$-2<C_0<-1$.
\end{enumerate}
These three scenarios are indicated in Fig.~\ref{F:scenarios}.

Consider the parameter range (iii) given by $B_0>0$ and $-2<C_0<-1$ which is plotted in the blue dash-dotted curves in Fig.~\ref{F:scenarios}: in 
the Einstein frame the Brans~IV solution describes a 
wormhole while in the Jordan frame, for the same parameter 
range, it describes a horizonless geometry. The reason for 
this change introduced by the conformal transformation is   
discussed in Sec.~\ref{sec:4}. 
The Ricci, Ricci squared, and Kretschmann scalars in the 
Einstein frame are
\begin{eqnarray}
\tilde{{\cal R}} &=& 
\frac{ 2\,\mbox{e}^{\frac{-2 \left( C_0+2 \right)}{B_0 
\varrho} } }{\varrho^2}  =\frac{2}{\tilde{R}^2}  \,,\\
&&\nonumber\\
\tilde{R}_{ab} \tilde{R}^{ab} &=& 
\frac{1}{2} \, \tilde{R}_{abcd} \tilde{R}^{abcd}=
\frac{ 2\,\mbox{e}^{\frac{-2 \left( C_0+2 \right)}{B_0 
\varrho} 
} }{\varrho^4} \,,
\end{eqnarray}
and are regular in the parameter range 
$B_0>0$ and $-2<C_0<-1$ since the coordinate $\varrho$ does 
not go below the minimum value $ \tilde{\varrho}_0$.

\section{The general problem}
\label{sec:3}

Let us consider now the general problem: how can it happen 
that a geometry without horizons, when 
conformally transformed, becomes a black hole or a 
wormhole? Surely something so physically dramatic as a 
horizon does not depend crucially on an overall factor in 
the 
metric! The answer is that there is no straightforward  
relation between (apparent) horizons in one geometry and 
(apparent) horizons of the conformally transformed metric, 
or even between the numbers of horizons present. 
Null geodesics, the causal structure of 
spacetime and event horizons (which are null surfaces)  are 
conformally invariant, but apparent and trapping 
horizons are spacelike or timelike surfaces 
and therefore they are affected by a conformal 
transformation.

Let us restrict, for simplicity, to spherically 
symmetric metrics, the line element of which can 
generically be written as 
\be\label{metric}
ds^2 =-A(t,R) dt^2 +B(t,R) dR^2 +R^2 d\Omega_{(2)}^2 \,.
\ee

\noindent This line element  is not 
required to be stationary and, clearly, $R$ is the areal 
radius. The apparent horizons, if they exist, are located 
by the equation 
$g^{RR}=1/B(t,R)=0$ (which may or may not have roots). 
In addition, in order to have a black hole and not  
a wormhole it is required that $g^{RR}$ have single roots, 
that is, $\partial g^{RR}/\partial R \neq 0$ where $g^{RR}$ 
vanishes.  

The 
conformally transformed geometry is given by
\be\label{metrictilde}
d\tilde{s}^2 =-\Omega^2 Adt^2 +\Omega^2 B dR^2 
+ \tilde{R}^2 d\Omega_{(2)}^2 \,,
\ee
where $\Omega=\Omega(t, R)$ to preserve the spherical 
symmetry and $\tilde{R}=\Omega(t,R) R$ is the areal radius 
in the conformally transformed world. 

One rewrites the ``new'' metric~(\ref{metrictilde}) 
in the form~(\ref{metric}) as
\be
d\tilde{s}^2 =-\tilde{A}( \tilde{t},\tilde{R}) d\tilde{t}^2 
+\tilde{B}( \tilde{t}, \tilde{R}) d\tilde{R}^2 +\tilde{R}^2 
d\Omega_{(2)}^2 \,, \label{czz}
\ee
where $\tilde{t}$ is a suitable time coordinate and 
$\tilde{R}$ is the areal radius in the Einstein frame. The 
form 
of the functions $\tilde{A}( \tilde{t},\tilde{R})$ and 
$ \tilde{B}( \tilde{t}, \tilde{R}) $ was 
computed in Ref.~\cite{ValerioEnzo2013}, which we follow 
here.

The use of the relation 
\be\label{12}
dR=\frac{d\tilde{R} - \Omega_{,t}Rdt}{ \Omega_{,R}R+\Omega} 
\ee
in the line element (\ref{metrictilde}) provides  
\begin{eqnarray}
d\tilde{s}^2 &=& -\left[ \Omega^2A 
-\frac{ \Omega_{,t}^2R^2\Omega^2 B}{ 
\left( \Omega_{,R}R+\Omega \right)^2}\right] dt^2 \nonumber\\
&&\nonumber\\ 
&\, & +\frac{ \Omega^2B}{ \left( \Omega_{,R}R+\Omega \right)^2 }\, 
d\tilde{R}^2 \nonumber\\
&&\nonumber\\
&\, & -\, \frac{2\Omega^2 \Omega_{,t}BR}{ 
\left( \Omega_{,R}R+\Omega \right)^2}\, dtd\tilde{R} 
+\tilde{R}^2 d\Omega_{(2)}^2 \,. \label{11ter}
\end{eqnarray} 
The $dt \, d\tilde{R}$ cross-term can be removed by 
the use of a new time coordinate $\tilde{t}(t, R)$ which 
satisfies 
\be \label{13}
d\tilde{t}=\frac{1}{F} \left( dt +\beta d R \right) 
\,,
\ee
with $\beta\left( t, R \right) $ an unknown function to 
be determined and $F\left(t, R \right)$ an 
integrating factor which necessarily obeys the 
differential relation 
\be\label{eqforF}
\frac{\partial}{\partial R}  \left( \frac{1}{F} 
\right) = 
\frac{\partial}{\partial t} \left( \frac{\beta}{F} \right) 
\ee 
which ensures that $d\tilde{t}$ is an exact differential.  
Substitution of $dt=Fd\tilde{t}-\beta d\tilde{R}$ into the 
line element yields
\begin{widetext}
\begin{eqnarray}
d\tilde{s}^2 &=&  -\left[ \Omega^2A -\frac{ 
\Omega_{,t}^2R^2\Omega^2 B}{ 
\left( \Omega_{,R}R+\Omega \right)^2}\right] F^2 
d\tilde{t}^2  +
\left\{ -\beta^2 \left[ \Omega^2A -\frac{ \Omega_{,t}^2R^2 
\Omega^2 B}{ \left( \Omega_{,R}R+\Omega \right)^2}\right] 
+\frac{ \Omega^2 B}{ \left( \Omega_{,R}R +\Omega \right)^2 
} \right.\nonumber\\
&&\nonumber\\
&\, & \left. 
+\frac{2 \beta \Omega_{,t}\Omega^2 BR}{ \left(  
\Omega_{,R}R+\Omega \right)^2}\right\} d\tilde{R}^2 
 +2F \left\{ \beta \left[ \Omega^2A -\frac{ \Omega_{,t}^2 
R^2\Omega^2 B}{ \left( \Omega_{,R}R+\Omega 
\right)^2}\right] - \frac{ 
\Omega_{,t}\Omega^2 R B}{ \left( \Omega_{,R}R +\Omega 
\right)^2}  \right\} d\tilde{t} \, d\tilde{R} +\tilde{R}^2 
d\Omega_{(2)}^2 \,. 
\end{eqnarray} 
We now fix the function $\beta$ as 
\be\label{16} 
\beta \left( t, R \right) = 
\frac{ \Omega_{,t} \Omega^2 BR}{ \left[ \Omega^2A -\frac{ 
\Omega_{,t}^2R^2\Omega^2 B}{ \left( \Omega_{,R}R+\Omega 
\right)^2}\right] \left( \Omega_{,R}R+\Omega \right)^2} 
\ee 
and the line element is diagonalized:  
\begin{eqnarray} 
d\tilde{s}^2 &=& 
-\left[ \Omega^2A -\frac{ \Omega_{,t}^2\Omega^2 BR^2}{ \left( 
\Omega_{,R}R+\Omega \right)^2}\right] F^2 d\tilde{t}^2 
+ \frac{ \Omega^2 B}{ 
\left( \Omega_{,R}R+\Omega \right)^2 }  
 \cdot \left\{ 1  + \frac{ \Omega_{,t}^2 
\Omega^2 BR^2}{ 
\left( \Omega_{,R}R+\Omega \right)^2  \left[ \Omega^2 
A -\frac{ \Omega_{,t}^2\Omega^2 BR^2}{ \left(  
\Omega_{,R}R+\Omega  \right)^2 } \right]} \right\} 
d\tilde{R}^2 \nonumber\\
&&\nonumber\\
&\,& +\tilde{R}^2 d\Omega_{(2)}^2 
\,.\label{17} 
\end{eqnarray}
\end{widetext}
By comparing eqs.~(\ref{17}) and~(\ref{czz}) one obtains 
\begin{eqnarray} 
\tilde{A} &=& \left[ 
\Omega^2A -\frac{ \Omega_{,t}^2\Omega^2 BR^2}{ \left( 
\Omega_{,R}R+\Omega \right)^2}\right] F^2  \,,\label{18}\\ 
&&\nonumber\\ 
\tilde{B} &=& \left.\frac{B\Omega^2}{ \left( \Omega_{,R}R+\Omega 
\right)^2}  \right\{ 1 \nonumber\\
&&\nonumber\\
&\, & \left. + \frac{ 
\Omega_{,t}^2 \Omega^2 BR^2}{ 
\left(  \Omega_{,R}R+\Omega \right)^2 \left[ \Omega^2 A 
-\frac{ 
\Omega_{,t}^2 \Omega^2 BR^2}{ \left( \Omega_{,R}R+\Omega 
\right)^2 } 
\right]} \right\} \,. \label{19} 
\end{eqnarray} 
the equation locating the apparent horizons of the tilded 
geometry is now $1/\tilde{B}=0$. Restricting to static, 
spherically symmetric metrics (such as those of 
Sec.~\ref{sec:2}), this equation reduces to 
\be\label{newAHs}
\frac{ \left( \Omega_{,R}R+\Omega \right)^2}{B\Omega^2} =0 
\,.
\ee
The roots of this equation are obtained by combining:
\begin{enumerate}
\item the roots of the ``old'' equation locating 
the apparent horizons $1/B=0$ (if they exist and they fall 
in the physical spacetime region).

\item The roots of $\Omega \rightarrow +\infty$; these are 
usually discarded because the conformal transformation 
becomes ill-defined there, but sometimes one can perform 
a conformal transformation and then extend the new 
solution thus obtained to spacetime regions in which the 
original conformal transformation is not defined 
(``conformal continuation'' \cite{Bronnikov}).

\item The roots of the equation $ \Omega_{,R}R+\Omega=0$. 
These roots (if they exist and lie in the physical 
spacetime region) are always double roots of 
eq.~(\ref{newAHs}).
\end{enumerate}

Suppose that the non-tilded geometry $g_{ab}$ had 
apparent horizons 
located by the roots of $g^{RR}=1/B=0$; the apparent 
horizons of its conformal cousin will correspond to the 
roots of $\tilde{g}^{\tilde{R}\tilde{R}}=0$. These roots 
will include the old ones (expressed in terms of the 
rescaled radius $\tilde{R}$) if they fall in the physical 
spacetime region. In general, the ``old'' apparent horizons 
could belong to a region of the new geometry separated by 
the ``physical'' region of the ``new'' geometry by a 
singularity. In this way, old apparent horizons could 
disappear from the conformally transformed geometry. New 
ones can appear because the rather complicated equation 
$1/\tilde{B}=0$ can in principle have new roots in addition 
to those of the simpler equation $1/B=0$.

It is also possible that black hole apparent horizons of 
the ``old'' metric which are roots of $g^{RR}=0$ with 
$\partial g^{RR}/\partial R \neq 0$ become apparent 
horizons of the ``new'' metric  corresponding to 
$\tilde{g}^{\tilde{R}\tilde{R}}=0$ but with 
$\partial g^{RR}/\partial R $ vanishing there: in this case 
the conformal cousin of the black hole apparent horizons 
are located at a wormhole throat. Armed with this 
understanding, let us revisit the 
example of Sec.~\ref{sec:2}.

\section{Revisiting Brans class~IV metrics}
\label{sec:4}

Let us examine the equation $ \Omega_{,R}R+\Omega=0$ 
which introduces possible new roots following the conformal 
transformation to the Einstein frame of Brans class~IV 
geometries. Since 
\be \label{etc}
\Omega=\sqrt{\phi}=\sqrt{\phi_0} \, 
\mbox{e}^{-\frac{C_0}{2B_0\varrho}} \,,
\ee
the expression of the areal radius~(\ref{arealradius}) 
gives
\be
\Omega_{,R} \equiv \frac{d\Omega}{dR}= 
\frac{d\Omega}{d\varrho} \, \frac{d\varrho}{dR}=
\sqrt{\phi_0} \, \mbox{e}^{-\frac{C_0}{2B_0\varrho} } 
\left[ 
\frac{C_0}{2B_0\left( \varrho-\varrho_0 \right)}+1 \right] 
\,,
\ee
hence eq.~(\ref{newAHs}) has the only root 
\be
\varrho = \varrho_0-\frac{C_0}{2B_0}= 
\frac{C_0+1}{B_0}-\frac{C_0}{2B_0} =
\frac{C_0+2}{2B_0} \,,
\ee
which is exactly $ \tilde{\varrho}_0$ defined by 
eq.~(\ref{rho1}). 
As 
discussed above, this is a {\em double} root of 
eq.~(\ref{newAHs}) locating the apparent horizons and 
signals a wormhole throat. 
Therefore, for Einstein frame Brans~IV geometries the 
apparent horizons originate as roots of the 
equation $\Omega_{, R}R +\Omega=0$  introduced by the 
conformal 
transformation where, in the Jordan frame, there were no 
apparent horizons ({\em i.e.}, no roots of $1/B=0$).

\section{Conclusions}
\label{sec:5}

We have clarified the nature of Brans class~IV solutions of 
Brans-Dicke theory. These geometries do not contain black 
holes, as is sometimes implied in the literature, but they 
describe wormholes or naked singularities. Now, 
asymptotically flat black holes in scalar-tensor gravity 
are the same as in General Relativity, according to well 
known no-hair theorems \cite{Hawking72, 
SotiriouFaraoniPRL, Romano}. The only exceptions are 
``maverick'' 
black holes which are either unstable, or in which the 
scalar field diverges somewhere (usually on the horizon) or 
vanishes there (in which case the gravitational coupling 
strength diverges) \cite{SotiriouFaraoniPRL}. If Brans~IV 
solutions (which have regular nonvanishing scalar $\phi$) 
did indeed describe black holes, they would violate the 
no-hair theorems, but we have shown them to describe 
wormholes or naked singularities in all their parameter 
range instead.

Turning to the general problem of the changing nature of  
a spacetime geometry subject to conformal mappings, it 
seems odd that 
a solution of scalar-tensor gravity hosts 
no horizons and describes a naked singularity in one 
conformal frame, while it admits a horizon and describes a 
wormhole in another conformal frame. This change under 
conformal mappings is even more surprising if one adopts 
the widespread point of view (which goes back to Dicke 
\cite{Dicke}) that different conformal frames are merely 
different representations of the same physical theory, with 
the condition that units of length, time, and mass change 
in the Einstein frame while they are fixed in 
the Jordan frame \cite{Dicke}. According to Dicke, 
under a conformal rescaling of the metric $g_{ab} 
\rightarrow \Omega^2 g_{ab}$, the units of length and 
time scale as $\Omega$, 
while the unit of  mass scales as $\Omega^{-1}$ and 
derived units scale 
according to their dimensions \cite{Dicke}. The physics 
in the two frames is the same once the rescaling of units 
is taken into account because all that is measured in a  
physical experiment (and all that matters from an  
operational point of view) is the ratio of a quantity to 
its unit. For example, both the mass of a particle  and 
its unit scale as 
$\Omega^{-1}$ in the Einstein frame and their ratio 
stays constant \cite{Dicke}. While it is hard to disagree 
with Dicke's view in principle, 
matters are more complicated in practice. Dicke's argument 
is easy to follow for test particles: Jordan frame 
timelike geodesics are mapped into non-geodesic curves in 
the Einstein frame because of the variation of particle 
masses in this frame\footnote{Null geodesics, of course, 
remain 
null geodesics under conformal rescalings as they suffer  
only an irrelevant change of parametrization 
\cite{Waldbook}.} \cite{Dicke}. However, the apparent 
horizons located by the roots of the equation 
$\nabla_c R\nabla^c R=0$ in spherical symmetry (or by the  
more complicated definition using the expansions of ingoing 
and outgoing null geodesic congruences in the absence of 
spherical symmetry \cite{NielsenVisser, AbreuVisser, 
VFhorizons}) are not related in any simple way to test 
particles. In spherical symmetry, apparent horizons are 
related to the Misner-Sharp-Hernandez mass $M_\text{MSH}$ 
of spacetime \cite{MSH}, which is defined by the equation
\be
1-\frac{2M_\text{MSH}}{R}=  \nabla_cR\nabla^cR=0 \,,
\ee 
which in turn gives $ R_\text{AH}=M_\text{AH}/2$ at the 
apparent horizons. It is 
well 
known that the Hawking-Hayward quasilocal energy 
\cite{Hawking, Hayward} (defined in General 
Relativity without restrictions of symmetry or 
asymptotic flatness) reduces to the Misner-Sharp-Hernandez 
mass in spherical symmetry \cite{Haywardspherical}. The 
Hawking-Hayward mass and its special case, the 
Misner-Sharp-Hernandez mass, are complicated 
integrals involving several physical quantities constructed 
with the ingoing and outgoing null geodesics through a  
compact, spacelike, orientable 2-surface 
and their gradients \cite{Hawking, Hayward}. It is not 
surprising, therefore, that the quasi-local energy does not 
transform simply as $\Omega^{-1}$ (that is, as the mass of 
a test particle), as expected naively from  
Dicke's dimensional considerations \cite{Dicke}. (The 
transformation property of the Misner-Sharp-Hernandez mass 
in spherical symmetry was worked out in 
\cite{ValerioEnzo2013} and that of the full Hawking-Hayward 
mass in \cite{hhconfo}.) Since apparent horizons are 
intimately related to a complicated quantity such as the 
quasilocal mass, it is not surprising that they change in 
non-trivial ways under conformal rescalings of the 
spacetime metric. While, from a fundamental point of view, 
lengths and times may just scale as $\Omega$ and masses as 
$\Omega^{-1}$, the transformation properties of composite 
objects and more complicated physical quantities may be 
much harder to predict and even counterintuitive. 
Therefore, although the two conformal frames may ultimately 
be physically equivalent at the classical microscopic 
level, in practice this equivalence can  be well hidden and 
is not manifest when considering  horizons and  their 
conformal rescalings.

\begin{acknowledgments} 
{\footnotesize  
VF and AP are grateful to Alex Nielsen for a discussion. 
This work is supported by Bishop's University and by the 
Natural Sciences and Engineering Research Council of 
Canada.} \end{acknowledgments}


\begin{thebibliography}{99}

\bibitem{Waldbook} R.M. Wald, {\em General Relativity} 
(Chicago University Press, Chicago, 1984).

\bibitem{BD61} C.H. Brans and R.H. Dicke, {\em Phys. Rev.} 
{\bf 124}, 925 (1961).

\bibitem{FujiiMaeda} Y. Fujii and K. Maeda, {\em The 
Scalar-Tensor Theory of Gravitation} (Cambridge University 
Press, Cambridge, 2003).

\bibitem{mybook} V. Faraoni, {\em Cosmology in Scalar
Tensor Gravity} (Kluwer Academic, Dordrecht, 2004).

\bibitem{CapozzielloFaraonibook} S. Capozziello and V. 
Faraoni, {\em Beyond Einstein Gravity} (Springer, New 
York, 2010).

\bibitem{reviews} T.P. Sotiriou and V. Faraoni, {\em Rev. 
Mod. Phys.} {\bf 82}, 451 (2010); 
A. De Felice and S. Tsujikawa, {\em Living Rev. Relativity} 
{\bf 13}, 3 (2010); S. Nojiri and S.D. Odintsov, {\em Phys. 
Repts.} {\bf 505}, 59 (2011).

\bibitem{noted1} K.K. Nandi, B. Bhattacharjee, S.M.K. 
Alam, and J. Evans, {\em Phys. Rev. D} {\bf  57}, 823 
(1998).

\bibitem{noted2} P.E. Bloomfield, {\em Phys. Rev. D} {\bf  
59}, 088501 (1999).

\bibitem{noted3} K.A. Bronnikov, M. S. Chernakova, J.C. 
Fabris, N. Pinto-Neto, and M.E. Rodrigues, {\em Int. J. 
Mod. Phys. D} {\bf 17}, 25 (2008).

\bibitem{Alex} V. Faraoni and A.B. Nielsen, {\em  
Class. Quantum Grav.} {\bf 28}, 175008 (2011)


\bibitem{Brans62} C.H. Brans, {\em Phys. Rev.} {\bf 125}, 
2194 (1962).

\bibitem{BhadraNandi2001} A. Bhadra and K.K. Nandi, {\em 
Mod. Phys. Lett.} {\bf 16}, 2079 (2001).

\bibitem{BhadraSarkarGRG2005} A. Bhadra and K. Sarkar, {\em 
Gen. Rel. Gravit.} {\bf 37}, 2189 (2005).

\bibitem{Weinberg} S. Weinberg, {\em Gravitation 
and Cosmology} (Wiley, New York, 1972).

\bibitem{RomeroBarros} C. Romero and A. Barros, {\em Phys. 
Lett. A} {\bf 173}, 243 (1993).

\bibitem{BanerjeeSen} N. Banerjee and S. Sen, {\em 
Phys. Rev. D} {\bf 56}, 1334 (1997).

\bibitem{VFlimit} V. Faraoni, {\em Phys. Lett. A} {\bf 
245}, 26 (1998); {\em Phys. Rev. D} {\bf 59}, 084021 
(1999).

\bibitem{BhadraNandilimit} A. Bhadra and K.K. Nandi, {\em 
Phys. Rev. D} {\bf 64}, 087501 (2001).

\bibitem{NielsenVisser} A.B. Nielsen and M. Visser, {\em 
Class. Quantum Grav.} {\bf 23}, 4637 (2006).

\bibitem{AbreuVisser} G. Abreu and M. Visser, {\em Phys. 
Rev. D} {\bf 82}, 044027 (2010).

\bibitem{VFhorizons} V. Faraoni, {\em Cosmological and 
Black hole Apparent Horizons} (Springer, New York, 2015).

\bibitem{Buchdhal} H. Weyl, {\em Ann. Phys. (Paris)} {\bf 
54} 117 (1917); H.A. Buchdahl, {\em Int. J. Theor. Phys.} 
{\bf 24}, 731 (1985).

\bibitem{Vanzoetal2012} L. Vanzo, S. Zerbini, and V. 
Faraoni, {\em Phys. Rev. D} {\bf 86}, 084031 (2012).

\bibitem{CampanelliLousto} M. Campanelli and C. Lousto, 
{\em Int. J. Mod. Phys. D} {\bf 02}, 451 (1993); C. Lousto 
and M. Campanelli, in {\em The Origin of Structure in the 
Universe}, Pont d'Oye, Belgium, 1992, edited by E. Gunzig 
and P. Nardone (Kluwer Academic, Dordrecht, 1993), p. 123.

\bibitem{Fisher} I.Z. Fisher, {\em Zh. Eksp. Teor. Fiz.} 
{\bf 18},  636 (1948) [arXiv:gr-qc/9911008]; O. Bergman and 
R. Leipnik, {\em Phys. Rev.} {\bf 107}, 1157 (1957); A.I. 
Janis, E.T. Newman, and J. Winicour, {\em Phys. Rev. Lett.} 
{\bf 20}, 878 (1968); H.A. Buchdahl, {\em Int. J. Theor. 
Phys.} {\bf 6}, 407 (1972); M. Wyman, {\em Phys. Rev. D} 
{\bf 24}, 839 (1981).

\bibitem{Cognolaetal11} G. Cognola, O. Gorbunova, L. 
Sebastiani, and S. Zerbini, {\em Phys. Rev. D} {\bf 
84}, 023515 (2011).

\bibitem{ValerioEnzo2013} V. Faraoni and V. Vitagliano, 
{\em Phys. Rev. D} {\bf 89}, 064015 (2014).

\bibitem{Bronnikov} K.A. Bronnikov and M.S. Chernakova, 
{\em Gravitation Cosmol.} {\bf 11}, 305 (2005).

\bibitem{Hawking72} S.W. Hawking, {\em Commun. Math. Phys.} 
{\bf 25}, 167 (1972).

\bibitem{SotiriouFaraoniPRL} T.P. Sotiriou and V. Faraoni, 
{\em Phys. Rev. Lett.} {\bf 108}, 081103 (2012).

\bibitem{Romano} S. Bhattacharya, K.F. Dialektopoulos, A.E. 
Romano, and T.T. Tomaras, arXiv:1505.02375.

\bibitem{Dicke} R.H. Dicke, {\em Phys. Rev.} {\bf 125}, 
2163 (1962).

\bibitem{MSH} C.W. Misner and D.H. Sharp, {\em Phys. Rev.} 
{\bf 136}, B571 (1964); 
W.C. Hernandez and C.W. Misner, {\em Astrophys. J.} {\bf 
143}, 452 (1966).

\bibitem{Hawking} S. Hawking, {\em J. Math. Phys.} {\bf 9}, 
598 (1968).

\bibitem{Hayward} S.A. Hayward, {\em Phys. Rev. D} {\bf 
49}, 831 (1994).

\bibitem{Haywardspherical} S.A. Hayward, {\em Phys. Rev. D} 
{\bf 53}, 1938 (1996).

\bibitem{hhconfo} A. Prain, V. Vitagliano, V. Faraoni, and 
M. Lapierre-L\'eonard, arXiv:1501.02977.


\end{thebibliography}

\end{document}